\documentclass[18pt,preprint]{aastex}
\begin{document}
\title{Diffuse Ionized Gas in the Dwarf Irregular Galaxy DDO 53 }
\author{A.M. Hidalgo-G\'amez}
\affil{Escuela Superior de F\'{\i}sica y Matem\'aticas, IPN, U.P. Adolfo L\'opez Mateos, Mexico City, 
Mexico\thanks{On leave from Instituto de Astronom\'{\i}a-UNAM}}

\begin{abstract}
The spectral characteristics throughout the dwarf irregular galaxy DDO 53 are studied. The results are 
very similar to those for other irregular galaxies: high excitation and low values of the 
[S\,{\sc ii}]/H$\alpha$ ratio. The most likely ionization source is photon leakage from the 
classical H\,{\sc ii} regions, without any other source, although the interstellar medium of the 
galaxy is quite perturbed. Moreover, the physical conditions throughout the galaxy do not change 
very much because both the photon leakage percentage and the ionization temperature are very similar. 
In addition, the determined metal content for two H\,{\sc ii} regions indicates that DDO 53 is a 
low-metallicity galaxy.

\end{abstract}

\section{Introduction}

Diffuse ionized gas  (DIG) is defined as ionized gas with a very 
low density. It is normally localized 
outside classical  H\,{\sc ii} regions, in the interarm regions of spiral 
galaxies (Benvenuti et al. 1976), at high 
altitude over the Galactic plane (Rand et al. 1990), and in very low density 
regions inside the Milky Way (Reynolds 1983). 
The spectral characteristics of this gas (low excitation and high values of the 
[N\,{\sc ii}]/H$\alpha$ and 
[S\,{\sc ii}]/H$\alpha$ ratios) indicate the existence of extra sources of 
ionization besides photoionization, such 
as shocks and turbulence due to stellar winds and other disturbances of the gas. 

Such a gas can also be observed in irregular galaxies, at the outermost parts 
(Hunter \& Gallagher 1992) 
and in the ``inter-H\,{\sc ii} regions'' (e.g. Hidalgo-G\'amez 2006). The 
spectral characteristics of the DIG in the few 
irregular galaxies studied so far (Hidalgo-G\'amez 2005, 2006; 
Hidalgo-G\'amez \& Peimbert 2007) indicate 
that the ionization sources are very different from those in spiral galaxies. 
The DIG of most of these galaxies can be 
ionized by photon leakage from the H\,{\sc ii} regions, although the
kinematics of the interstellar medium (ISM) is 
disturbed in some  galaxies (GR 8; Hidalgo-G\'amez 2006). Only a few of the 
galaxies studied so far need an extra source of 
ionization (e.g. Wolf-Rayet stars for IC 10; Hidalgo-G\'amez 2005). Our 
goal is to study the influence of the 
perturbations of the interstellar medium on the ionization of the DIG, if 
any, and whether there is any difference for 
different metallicities, gas masses, or star formation rates. As irregular 
galaxies comprise a large range in these 
properties, a large sample of them is needed.

DDO 53 is a not-very-well-studied dwarf irregular galaxy of the M81 
group, but it has three properties which make it a very good candidate for the study 
of the DIG. First, it has a high gas content with a total gas mass  
of $2.45\times~10^8$ $M\odot$ (Hunter \& Elmegreen 2004) and a star formation 
rate of $5\times~10^{-3}$ $M\odot$ yr$^{-1}$. 
Second, it is located at only $3.53$ Mpc (Hidalgo-G\'amez \& Olofsson 1998); 
therefore, the DIG can be detected easily 
with medium-size telescopes and not very long integration times. Moreover, the 
spatial resolution is good enough to study 
the details of the DIG and differentiate it from the classical H\,{\sc ii} regions. 
Finally, H$\alpha$ images of the 
galaxy (Strobel et al. 1990) indicate the existence of  gas in between the 
classical H\,{\sc ii} regions. 

A description of the observations and the data reduction procedure is given in 
Section 2. Section 3 presents a brief 
description of the DIG along the slit positions. The ionization source of the DIG is 
presented in Section 4, while in 
Section 5 a brief discussion on the spectral characteristics of the DIG of this galaxy 
compared with previous 
galaxies is given, as well as a determination of the metallicity. Finally, conclusions 
are given in Section 6.

\section{Observations and data reduction}

The spectra of DDO 53 were obtained on 2002 March 12-14 with the 
2.1 m telescope of the Observatorio Astron\'omico Nacional at San Pedro 
M\'artir. The Boller \& Chivens spectrograph was used with a 
300 line mm$^{-1}$ grating blazed at 5000 \AA. The detector was a SITe3 CCD with 3\,$\mu$m
pixels in a $1024 \times 1024$ format. The slit length was $5{\arcmin}$ and the width was 170\,$\mu$m, 
subtending $\approx$2${\arcsec}$ on the sky, and yielding a spectral 
resolution of $7$ \AA. The pixel scale along the spatial direction was $1.05 ''$. The full spectral range 
observed was 3500-6900 \AA.
During the first two nights, strong wind and cirrus made the conditions not  
very photometric, while the third night the weather conditions improved, 
although the seeing was higher. Table 1 shows the log 
of observations. No correction for differential refraction was performed, 
as the air masses were, in all 
cases, smaller than $1.3$. The orientation of 
the slit was east-west with a total of eight slit positions.

The reduction of the data was performed with the MIDAS software 
package. Bias and sky twilight flat-fields were used for the calibration of 
the CCD response. Due to the strong differences in the sensitivity of the 
CCD based on the wavelength, the spectral range was divided into three parts: 
from 5200 to 6800 \AA~ (red), from 4000 to 5600 \AA ~(green), and 
from 3500 to 4800 \AA~ (blue). Unfortunately, due to the low efficiency 
of the spectrograph at the blue end, no spectral lines were detected there, 
and the blue end was therefore not used in the present investigation.  He-Ar lamps 
were used for the wavelength calibration. In  addition, the spectra were
corrected for geometric deflections. The spectra were corrected for 
atmospheric extinction using the San Pedro M\'artir tables (Schuster \& 
Parrao 2001). Several standard stars were 
observed each night in order to perform the flux calibration. The accuracy 
of the calibration was higher than 5$\%$ for all nights. The most 
difficult task was the sky subtraction, because removing the 
strong sky lines of the spectra also removes most of the 
low surface brightness emission. In order not to remove this emission, 
only a dozen rows at each edge of the image were used in the sky 
templates for each spectrum, although the emission from the galaxy 
covered only the central part of the slit. The number of rows, different
for each spectrum, was selected as a compromise between the correctness of the results
and the maintenance of a low surface brightness emission. This was done by 
visual inspection. The main problem is that the intensity of neither He\,{\sc i} 
at $\lambda$5875 
nor [O\,{\sc i}] $\lambda$6300\AA~can be measured because of the proximity of strong 
sky lines, 
but these lines are not very commonly seen
at low surface brightness. Moreover, sky subtraction which removed all the sky lines, 
has been performed for
other data sets with similar results to those presented here (see, e.g. 
Hidalgo-G\'amez 2006). In any case, special care was taken in the analysis  of this 
data.

This set of data, fully reduced and calibrated, was divided into a number 
of spectra, covering five rows each. This value matches the seeing conditions 
during the observations and corresponds to $90$ pc at a distance of $3.53$ Mpc 
(Hidalgo-G\'amez \& Olofsson 1998). A total of $86$ one-dimensional spectra were 
studied, covering 
eight slit positions throughout the galaxy. These spectra are referred to as the
``1D spectra''. 

The next step is to correct these spectra for absorption and extinction. The usual 
procedure in the absorption correction is to increase the equivalent width of H$\beta$ 
by 2\AA~ (e.g. McCall et al. 1985). A more accurate absorption correction is to follow 
Olofsson (1995), who calculated the absorption equivalent width for each 
Balmer line using different initial mass functions, metallicities, and ages of the star 
formation event. 
All these procedures were optimized for H\,{\sc ii} regions but not for low surface 
brightness emission, such as studied here. In order to check whether the absorption 
corrections proposed could be used in the DIG spectrum, we measured other Balmer 
lines in the DIG when present, made the correction to them, and 
checked whether  the values obtained matched the theoretical ones.
The most easily detected line after H$\alpha$ and H$\beta$ is H$\gamma$. This line is detected 
in some of the spectra. Following McCall et al's correction, only one out of nine of the 
H$\gamma$ intensities matched the theoretical one. The same was true when considering 
Olofsson's correction. Therefore, as there were no clear absorption features 
present in the spectra studied here, we preferred not to perform any absorption correction. 

The same discussion can be made for the  extinction correction. There are two reasons 
we decided to perform this correction. First, the wavelength range for most of the 
line ratios is small enough not to be affected by extinction. Second, one of the 
goals of the investigation is to study the distribution of the extinction inside the galaxy. 
This correction was performed using the  extinction coefficient, defined as 
$$C\beta = {1\over f(\lambda)} ln{I(H\alpha)/I(H\beta) \over 2.86}$$ where $2.86$ is the theoretical 
Balmer decrement at $10,000$ K for case B of recombination, $f(\lambda$) is the 
Whitford modified extinction law (Savage \& Mathis 1979), and $I(H\alpha)/I(H\beta)$ is 
the observed H$\alpha$/H$\beta$ ratio. The intensities of the lines H$\beta$, [O\,{\sc iii}] 
$\lambda$5007, H$\alpha$, [N\,{\sc ii}] $\lambda$6583, He\,{\sc i} $\lambda$6678,
[S\,{\sc ii}] $\lambda$6716, and [S\,{\sc ii}] $\lambda$6731 were  
measured when present, as well as the flux in the H$\alpha$ line. All the lines were 
measured twice, except when the two values differed by more than 50$\%$, normally for 
low signal-to-noise ratio (S/N) lines. For these, 
a third measurement with a different continuum level was obtained in order to find a 
more reliable 
value for the intensity.  With these lines the ratios [O\,{\sc iii}]/H$\beta$, 
[O\,{\sc iii}]/H$\alpha$, [N\,{\sc ii}]/H$\alpha$, He\,{\sc i}/H$\beta$ and 
[S\,{\sc ii}]/H$\alpha$ were 
obtained. 

Finally, we would like to comment on the uncertainties. Three different sources 
were considered: the uncertainties in the level of the spectral continuum 
with respect to the lines,  $\sigma_c$, those introduced by the 
reduction procedure,  (especially flat-fielding and flux 
calibrations), $\sigma_r$, uncertainties due to the extinction correction, 
$\sigma_e$.   The final uncertainty for each line was determined from 
$$\sigma = \sqrt {\sigma_c^2 + \sigma_r^2  + \sigma_e^2}$$ These 
uncertainties were measured for each line ratio at each spectrum in all the 
positions for each galaxy. With these values a total uncertainty for 
each line and each slit position can be determined. They are given in Table 2.

\section{Diffuse Ionized Gas}

As mentioned above, we are interested in the study of the spectral characteristics of the
DIG. They will be used to understand the ionization source of this gas 
and the nature of the H\,{\sc ii} regions. 

\begin{figure}
\centering
\includegraphics[width=12cm]{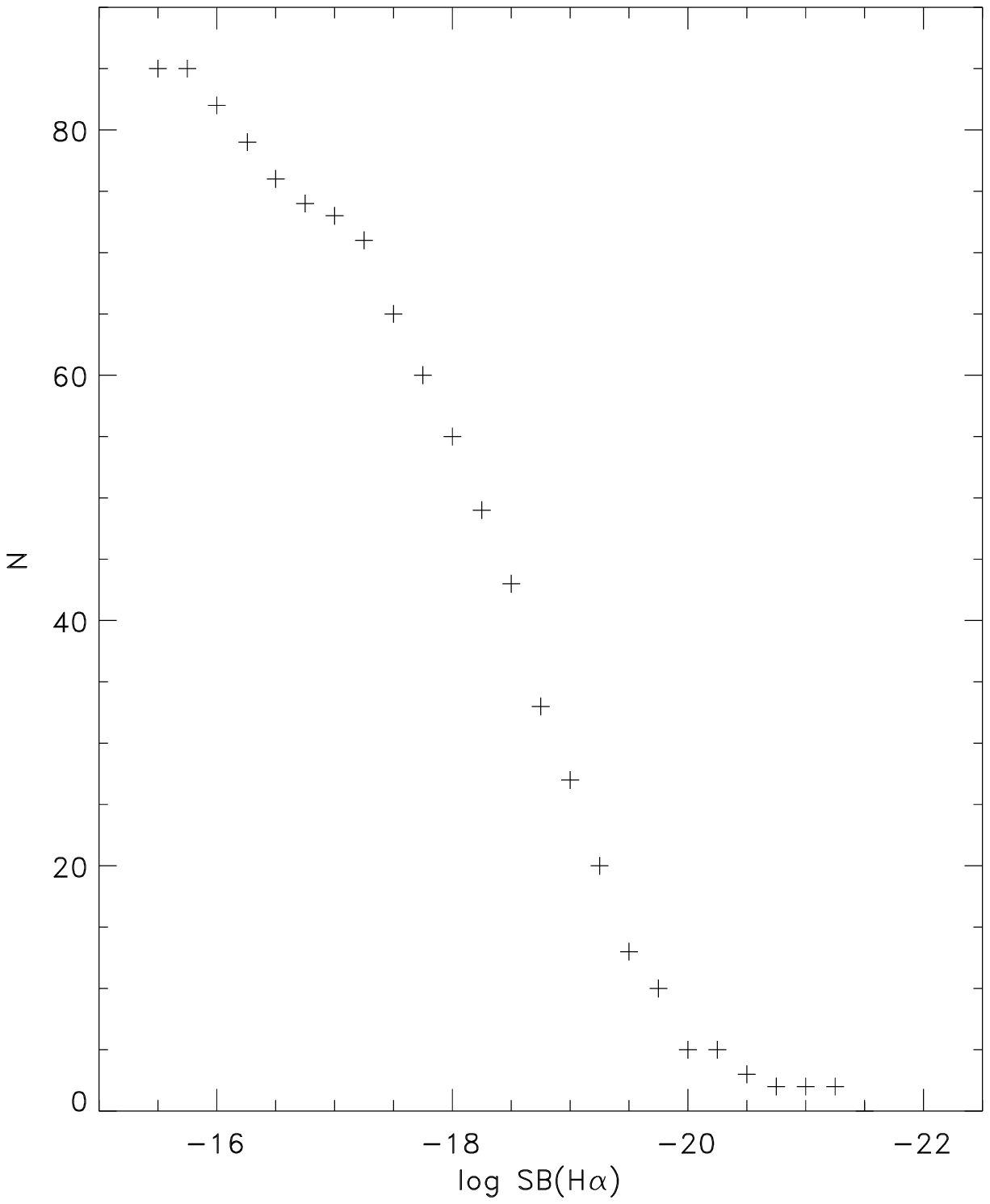}
\caption{ Cumulative function of SB(H$\alpha$) for DDO 53. Two break points are 
present. The one at - 20.5 gives the limiting value between the noise and the emission, 
while the other at - 17.3 divides the spectra between DIG and H\,{\sc ii} regions. }
\label{fig1}
\end{figure}

The best parameter to discriminate between ionized gas inside and outside the 
H\,{\sc ii} regions is the density. Unfortunately, density is one of the 
most difficult parameters to determine from 
long-slit spectroscopy data. The main problem is that in the low-density 
regime the differences in the [S\,{\sc ii}]$\lambda$6717/[S\,{\sc ii}]$\lambda$6731 ratio, 
which is 
the most common line ratio used in the density determination, between 
H\,{\sc ii} regions, with typical densities of $100$ cm$^{-3}$, and the DIG, 
with values of $10$ cm$^{-3}$, are about $0.1$ (see Figure 5.3 in Osterbrock 
1989). Such a difference is normally smaller than the uncertainties in the ratios 
themselves. In spiral
galaxies this can be overcome by studying the gas above the disc (e.g. Otte \& Dettmar 
1999) or in the inter-arm region (Benvenuti et al. 1976), with neither of them
clearly defined for irregular galaxies. Therefore, other parameters have to be 
used in the study of the DIG. In this sense the emission measure (EM), related to 
surface brightness in H$\alpha$ and the electronic temperature (Greenawalt 
et al. 1998), is a good choice. In the Milky Way there is a correlation between 
low EM and low density (R. Reynolds 2004, private communication). Therefore, 
the EM can be used to discriminate  
between DIG and H\,{\sc ii} regions when the density cannot be determined. The 
main problem is the lack of a unique value for the EM of the DIG in spiral galaxies. It varies from 
$80$ pc cm$^{-6}$ in the arms of spiral galaxies (Hoopes \& Walterbos 2003) 
to $2$ pc cm$^{-6}$ in the Milky Way (Reynolds 1989). Instead of choosing a 
value in this range, we decided to use another approach. The cumulative 
distribution function of the surface brightness in H$\alpha$ [hereafter 
SB(H$\alpha$)] is a smoothly increasing function  with a change in the 
slope when H\,{\sc ii} regions 
begin to dominate the light. This break point in  SB(H$\alpha$) can be 
considered as the limiting value between DIG and H\,{\sc ii} regions. This 
procedure has already been used for some galaxies (IC 10, 
Hidalgo-G\'amez 2005; NGC 6822, Hidalgo-G\'amez \& Peimbert 
2007; Gr 8 and ESO 245-G05, Hidalgo-G\'amez 2006). Figure ~\ref{fig1} shows 
the cumulative function of SB(H$\alpha$) obtained from the 1D spectra, 
and corrected for Galactic extinction following Schlegel et al. (1998). The break 
point is located at $5 \times 10^{-18}$ ergs s$^{-1}$ cm$^{-2}$ arcsec$^{-2}$, 
which corresponds to a flux of $5.45 \times 10^{-17}$ ergs s$^{-1}$ cm$^{-2}$. 
This value is very similar to the ones
found in GR 8 and ESO 245-G05 (Hidalgo-G\'amez 2006). In their study of the 
H\,{\sc ii} regions in DDO 53, Strobel et al. (1990) set the boundaries of 
their regions at a flux of $5 \times 10^{-17}$  ergs s$^{-1}$ cm$^{-2}$, which 
is very similar to the value found here. Another break point 
can be detected at $3.2 \times 10^{-21}$ ergs s$^{-1}$ cm$^{-2}$, 
arcsec$^{-2}$ which might be an indication of noise level. There are 
only five data points with SB(H$\alpha$) lower than this limit, and they are 
not going to be considered in this investigation.

Using this SB(H$\alpha$) from the 1D spectra, we can divide the data points 
into $58$ DIG and $13$ H\,{\sc ii} locations. Another $10$ spectra have no emission in 
any line. In addition to the 1D spectra, we can sum up all the rows with 
SB(H$\alpha$) lower (higher) than $5 \times 10^{-18}$ ergs s$^{-1}$ cm$^{-2}$ arcsec$^{-2}$ 
in order to create the so-called ``integrated spectra'' for the DIG (H\,{\sc ii} regions). 
A total of $26$ integrated spectra were created: $9$ H\,{\sc ii} regions 
and $17$ DIG locations.

\begin{figure}
\centering
\includegraphics[width=12cm]{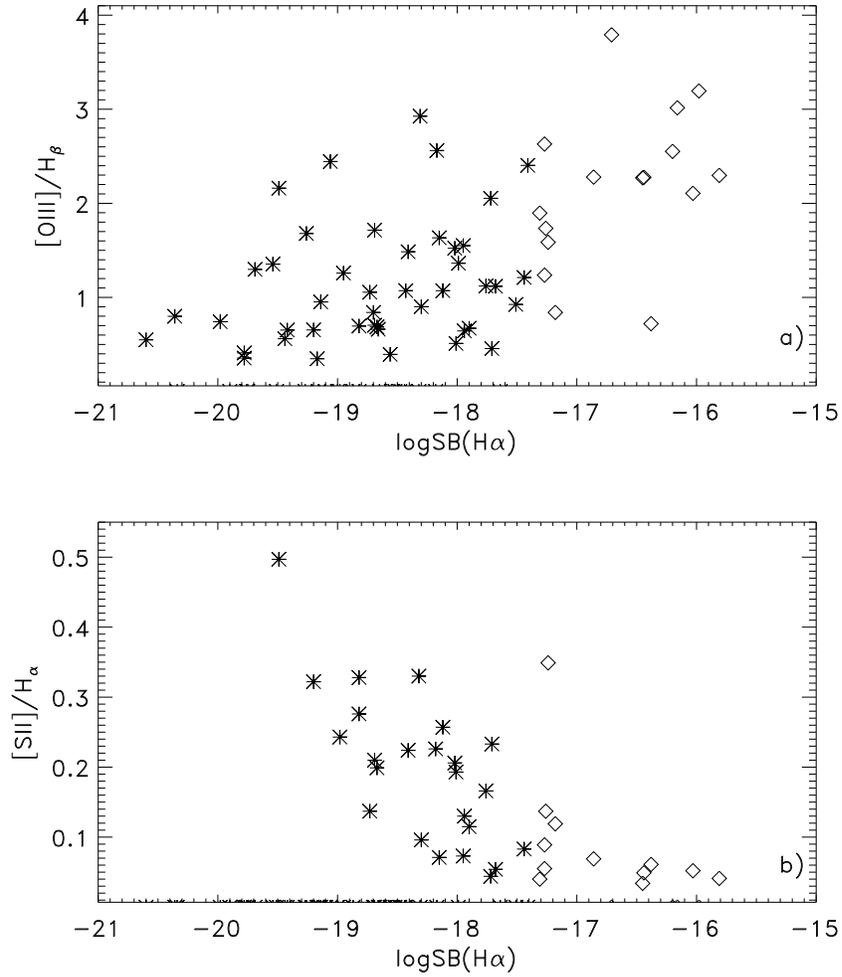}
\caption{Plot of log SB(H$\alpha$) vs. [O{\sc iii}]/H$\beta$ (top) and the log SB(H$\alpha$) vs. 
[S{\sc ii}]/H$\alpha$ (bottom) for DDO 53. The DIG locations are plotted as asterisk, and the 
H\,{\sc ii} regions as diamonds (see text for a definition). A smooth transition between 
the DIG and H\,{\sc ii} regions is clear from both plots, indicating that our definition of 
H\,{\sc ii} regions is correct. }
\label{fig2}
\end{figure}

The main caveat is that the cumulative function is a statistical approach, and 
therefore, the distinction between H\,{\sc ii} regions and DIG might be fuzzy, especially at 
low surface brightness or when the spatial averages  of the 1D spectra are large, as in 
NGC 6822, where the 1D spectra covered $50$ pixels (see Hidalgo-G\'amez \& Peimbert 
2007 for details). In order to test the goodness of this 
approach, we can study the relation between SB(H$\alpha$) and the line 
ratios, particularly the [O\,{\sc iii}]/H$\beta$ and [S\,{\sc ii}]/H$\alpha$ ratios. This is 
shown in Figure ~\ref{fig2}. As expected, there is a trend toward low [S\,{\sc ii}]/H$\alpha$ 
and high [O\,{\sc iii}]/H$\beta$ as SB(H$\alpha$) gets larger, with regression 
coefficients of $-0.45$ and $-0.3$, respectively. The transition between H\,{\sc ii} 
and DIG  locations is very smooth in the log [O\,{\sc iii}] versus SB(H$\alpha$) plot, 
although the dispersion is large. There are DIG locations with high excitation ($>2$) 
and H\,{\sc ii} regions with low excitation ($<1$). Two of the latter are located at the 
border of the H\,{\sc ii} regions,  and therefore, they might share the characteristics of both 
locations, while another two are small H\,{\sc ii} regions of only 1D spectra. 
Finally, the DIG with the highest excitation corresponds to slit f, where no H\,{\sc ii} 
regions were detected. The transition in Figure  ~\ref{fig2}b is also smooth, with a very 
steep correlation for the DIG locations. In any case, these figures indicate that the distinction 
made in this investigation is reasonable.

\subsection{Line Ratios along the Slits}

Once the distinction between the H\,{\sc ii} regions and the DIG locations is clear, 
the next step is to study some of the characteristics, such as the excitation, the 
distribution of SB(H$\alpha$), and the ratios [N\,{\sc ii}]/H$\alpha$ and 
[S\,{\sc ii}]/H$\alpha$ throughout the galaxy. This can be performed using the 
set of 1D spectra previously defined. Each of the 1D spectra covers 
$90$ pc, and therefore, the variations between them for the H\,{\sc ii} and DIG locations 
can be studied in detail.  

\begin{figure}
\centering
\includegraphics[width=10cm]{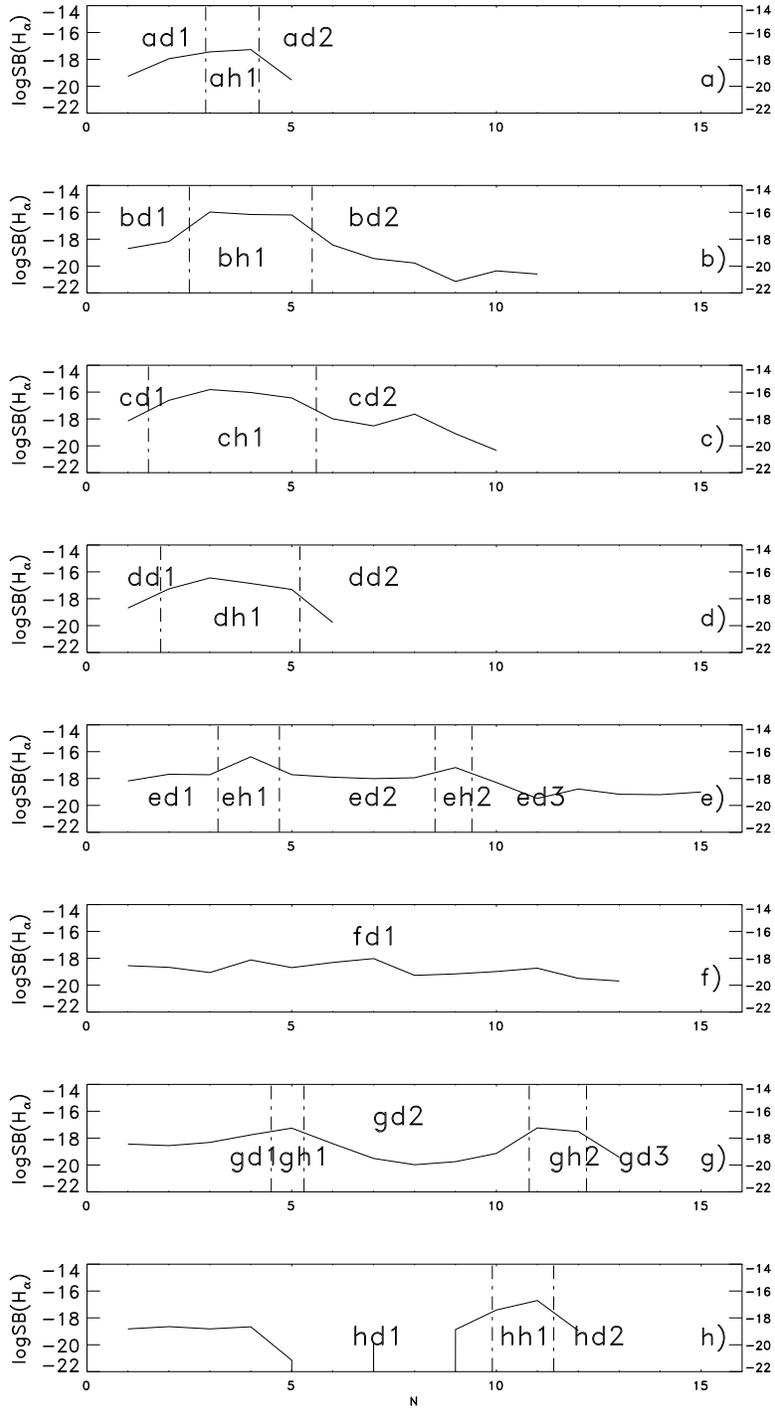}
\caption{SB(H$\alpha$) along the slits in DDO 53. Slit a corresponds 
to the northern part of the galaxy and slit h to the southern. West is to the left. 
The dot-dashed lines encompass those regions where the flux is higher than $5 \times 
10^{-18}$ ergs s$^{-1}$ cm$^{-2}$ arcsec$^{-2}$.  }
\label{fig3}
\end{figure}

Figure ~\ref{fig3} shows the variations of SB(H$\alpha$) along 
the slit, from north (at the top, slit a) to south (slit h) and from 
west (to the left) to east. All the slit positions have the same SB limits in order 
to see their differences throughout the galaxy. Those places with SB(H$\alpha$) higher than 
$5 \times 10^{-18}$ ergs s$^{-1}$ cm$^{-2}$  arcsec$^{-2}$ (H\,{\sc ii} regions) are contained between 
dot-dashed lines. The most extended H\,{\sc ii} regions are located at slits b, 
c and d, with sizes of $315$, $360$, and $315$ pc, respectively. As we move 
towards the south the H\,{\sc ii} regions get smaller and the DIG  
is more extended. In fact, in one slit position (f) there are no 
H\,{\sc ii} regions. When compared 
with an H$\alpha$ image (e.g. Figs. 2{\it a} and 2{\it b} in Strobel et al. 1990), the similarities are 
clear. The largest H\,{\sc ii} regions are located in the north of the 
galaxy, while towards the south, only small clumps of high intensity are found 
with a lot of ionized gas among them. The regions in this investigation can be identified 
as regions No $9$ (slit a), No $10$ at slit b, No $13$ at slits c and d, 
No $5$ and $11$ (e), No $8$ and $16$ (g) and No $17$ at slit h in 
Strobel et al. (1990). It has to be kept 
in mind that the slit positions are not next to each other, with spatial 
gaps between the slits. Moreover, the slits were offset towards the west when moving 
from north to south following the orientation of the galaxy.  It is interesting to 
note that there are no abrupt changes in SB(H$\alpha$) from the H\,{\sc ii} regions 
towards the DIG, as in NGC 6822 (see Fig. 3 in Hidalgo-G\'amez \& Peimbert 2007).

\begin{figure}
\centering
\includegraphics[width=10cm]{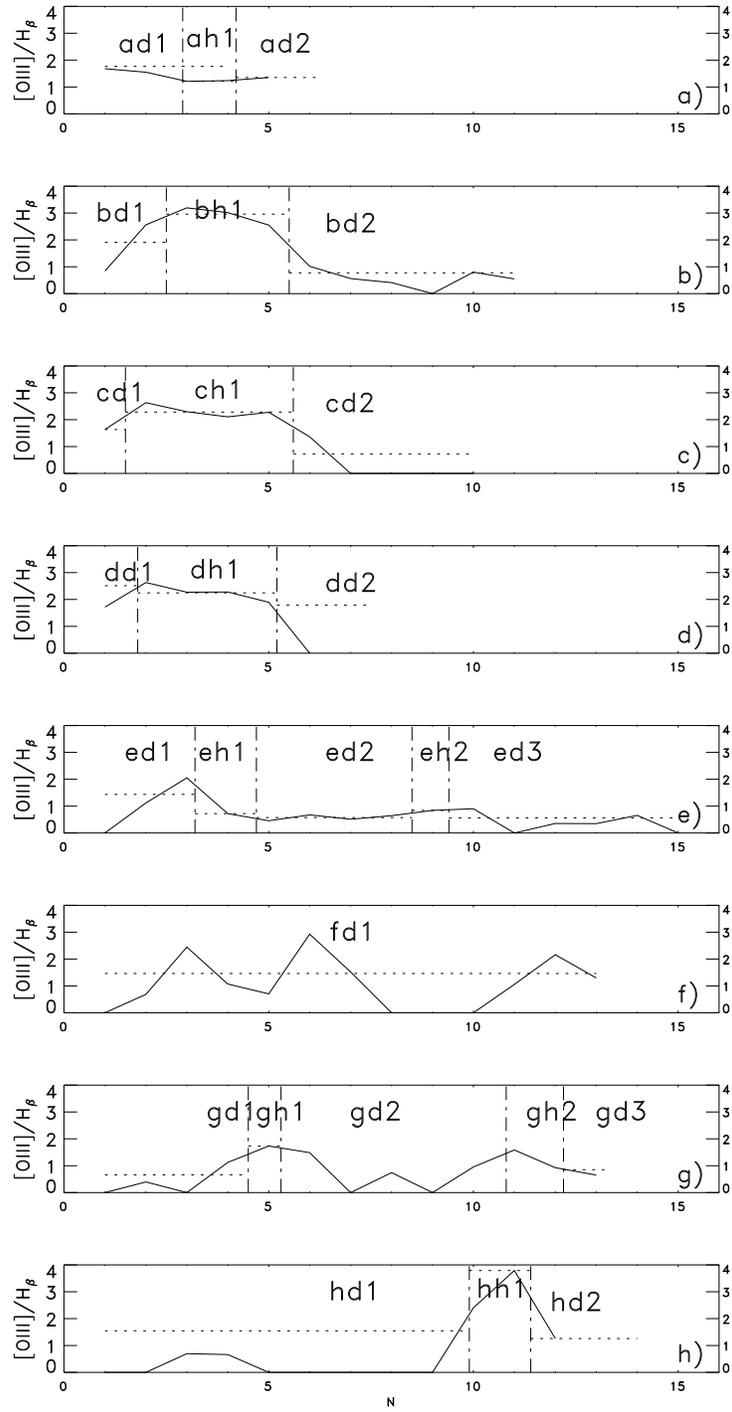}
\caption{[O\,{\sc iii}]/H$\beta$ along the slits in DDO 53. The symbols and 
orientation are as in Fig. ~\ref{fig3}. In addition to those, the dotted lines correspond 
to the values from the integrated spectra. }
\label{fig4}
\end{figure}

The [O\,{\sc iii}]/H$\beta$ ratio along the slits is shown in Figure ~\ref{fig4}. 
The orientation and the meaning of the lines are the same as in Figure ~\ref{fig3}. Along 
with the H\,{\sc ii} regions, the values from the integrated spectra are shown 
as dotted lines. No uncertainties are shown for the sake of clarity. Also, the limits 
in all the slit positions are the same in 
order to see any differences among them. The largest value of the excitation 
of $3.38\pm0.01$ is at slit h, corresponding to region No $17$, while 
slits b, c and d show [O\,{\sc iii}]/H$\beta$ between 
$2.5$ and $3$. They coincide with the largest and brightest H\,{\sc ii} 
regions of the galaxy. The largest value of the excitation is normally 
coincident with the largest SB except for slit e, where the maximum 
excitation is at the border of one of the H\,{\sc ii} regions. 
Moreover, the excitation inside the H\,{\sc ii} regions of slit e 
is very low ($\le 1$ in both cases) and very similar to the values 
outside them. On the contrary, the excitation is very high ($\ge 1$) 
for most of the DIG locations of slit f. Both slit g and h 
are normal, with the highest excitation inside the H\,{\sc ii} 
regions and a low value of the excitation outside them, in general. 
Concerning the DIG, it is well known that the [O\,{\sc iii}]/H$\beta$ 
ratio is lower than $1$ in the DIG of spiral galaxies (Rand 1998). This 
is not the situation here, since most of the values of this ratio are 
larger than $1$ at the DIG. In fact, three of the 1D spectra of 
slit f, where no proper H\,{\sc ii} regions are, have 
[O\,{\sc iii}]/H$\beta$ larger than $2$. Such large values of 
this ratio for DIG locations have been measured in other dwarf irregular  
galaxies, such as IC 10 (Hidalgo-G\'amez 2005), NGC 6822 (Hidalgo-G\'amez \& 
Peimbert 2007), and ESO 245-G05 (Hidalgo-G\'amez 2006). Another interesting fact 
to note is that only for slits b, c, g, and h is this ratio 
larger inside the H\,{\sc ii} regions than in the DIG locations, as 
expected. For the rest of the slits (a, d, and e), the 
value of the ratio in the DIG is smaller than or of the same order as inside the 
H\,{\sc ii} regions. For all of them, the [O\,{\sc iii}]/H$\beta$ ratio from the 1D spectra 
agrees with the values from the integrated spectra when the uncertainties are considered.

\begin{figure}
\centering
\includegraphics[width=10cm]{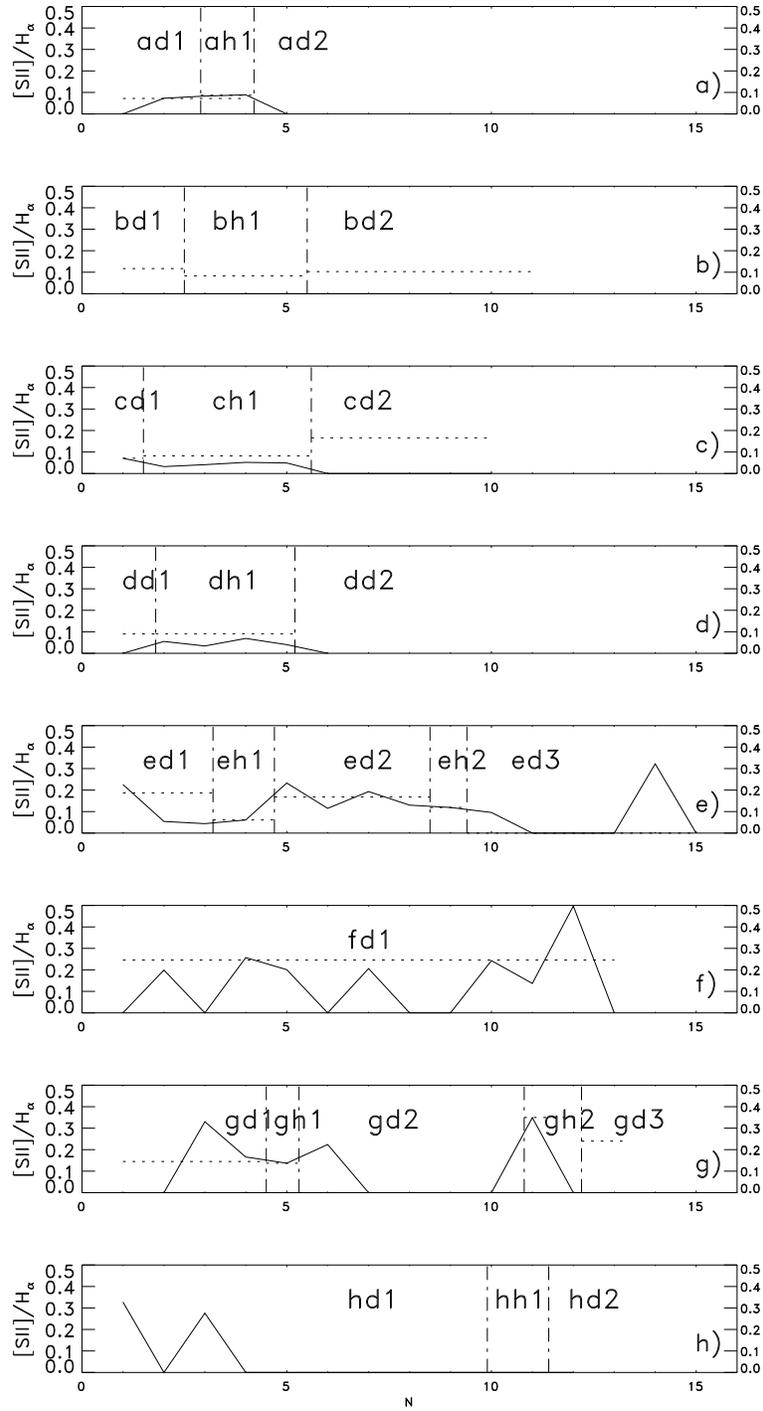}
\caption{[S\,{\sc ii}]/H$\alpha$ ratio along the slits in DDO 53. The symbols 
and orientation are as in Fig. ~\ref{fig4}. }
\label{fig6}
\end{figure}

Figure ~\ref{fig6} shows the [S\,{\sc ii}]/H$\alpha$ ratio along the slit positions. 
The orientation and the meaning of the lines are the same as in Figure ~\ref{fig3}, 
and the values are between $0$ and $0.5$ for all slit positions.  
Again, the average values are shown as dotted lines, when determined.  In 
previous studies an important difference in this ratio between the DIG and 
the H\,{\sc ii} locations has been detected for both spiral (NGC 55; Otte 
\& Dettmar 1999) and irregular galaxies (NGC 6822; Hidalgo-G\'amez \& Peimbert 
2007). It is very interesting to note that, in general, there is no such  
difference here, either in the  integrated spectra or in the 1D ones, with 
similar values of the [S\,{\sc ii}]/H$\alpha$ ratio in the DIG and inside the 
H\,{\sc ii} regions (e.g. slits a, c and d). For those 
slit positions where the differences are larger, they are not as large as in 
other irregular galaxies, such  as NGC 6822 (Hidalgo-G\'amez \& Peimbert 2007). 
In fact, a very large value of this ratio is detected inside 
the H\,{\sc ii} region  gh2 for both the 1D spectra and the 
integrated ones, which is quite unexpected.  Moreover, sulfur lines 
were not detected at any locations along slit b but, due to the larger S/N, 
a value is obtained from the average spectrum. On the contrary, 
[S\,{\sc ii}]/H$\alpha$ is measurable in two of the 1D spectra of slit 
h, but no value could be obtained for the average spectrum because 
the number of 1D spectra where this line is not detected was larger 
than the number of locations where it is detected. In this case, the noise 
from the non detections is larger than the signal.

The [S\,{\sc ii}]/H$\alpha$ ratio is important because  it might be a 
shock indicator (Dopita 1993). For low metallicities, intensities larger 
than $0.3$ indicate the existence of shocks (Stasi\'nska 1990) when the 
excitation is low. Such large values are not observed except for a total of 
four 1D spectra in slits e, f and g, and none 
of them has a high excitation as well.  Three of them are located in the 
eastern part of the galaxy. Begum et al. (2006), in their study of the 
interstellar medium of this galaxy, obtained velocity maps. They found 
out that the velocity gradient towards the south eastern part of the 
galaxy, which is receding from us, is larger than that in the northern 
approaching half of the galaxy. Apart from this, there is nothing else 
peculiar in the H\,{\sc i} and velocity maps. An increment in the 
[S\,{\sc ii}]/H$\alpha$ ratio towards the south is clearly seen in 
Figure ~\ref{fig6}, which might resemble the increase in the velocity 
gradient. In this sense, the lack of this ratio in the southernmost 
slit position, h, is even more intriguing. From this information 
it can be advanced that shock might not be important as an ionization 
source, as we discuss in section 4, in spite of the disturbed 
kinematics of the galaxy (Begum et al. 2006).

The [N\,{\sc ii}]/H$\alpha$ ratio is not shown here due to the small number of 
1D spectra where nitrogen was detected. Therefore, few conclusions 
could be obtained from this distribution. The few points detected in the 
northern slits have very low values of this ratio, while they increase 
when moving towards the south, with ratios of $0.3$ at slit e.  

\begin{figure}
\centering
\includegraphics[width=10cm]{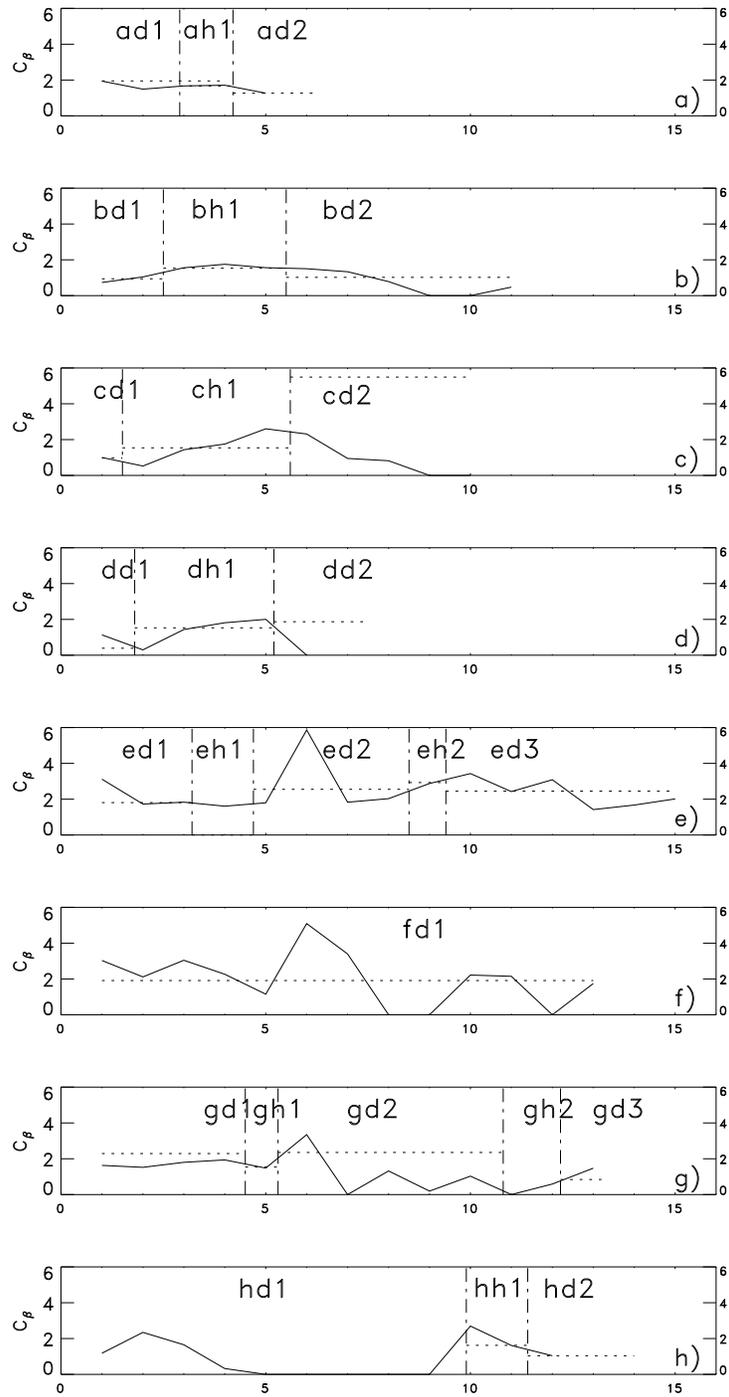}
\caption{Extinction coefficient along the slits in DDO 53. The symbols and orientation 
are as in Fig. ~\ref{fig4}. }
\label{fig7}
\end{figure}

Finally, Figure ~\ref{fig7} shows the extinction along the slits. It is 
parametrized with the extinction coefficient defined as  in Section 2.
The values are corrected for Galactic extinction 
following Schlegel et al. (1998). DDO 53 shows an irregular extinction 
distribution, not only along the slits but also between them. There are 
slits where the extinction seems to be very homogeneous (e.g.  a-d ), while slits 
e-h show (extended) clumps of it, as can be seen in the H$\alpha$ 
image from Strobel et al. (1990). This might resemble the fact that at this 
part of the galaxy the H\,{\sc ii} regions also have small sizes, while at 
the northern part the H\,{\sc ii} regions have larger sizes. This behaviour 
is very similar to that of other galaxies, such as GR 8 (Hidalgo-G\'amez 2006). As 
in IC 10 (Hidalgo-G\'amez 2005) and GR 8 (Hidalgo-G\'amez 2006), the larger 
extinction is not associated with the H\,{\sc ii} regions and vice versa. The 
maxima are located at e6 and f6 with values of $6 \pm 1$ and 
$5 \pm 1$, respectively. They are not associated with the two clumps observed 
in H\,{\sc i} (Begum et al. 2006), and they are mainly coincident with the
 northern regions, No $9$ and No $10$, and the south eastern ones No $17$ and $18$. 
In any case, it is interesting to note that most of the values of $C_{\beta}$ 
are larger than $1$. This is contrary to the idea that dwarf galaxies are almost 
dust-free (Lee et al. 2003) and in agreement with the conclusions from high 
spatial resolution investigations carried out in the Small Magellanic Cloud 
by Caplan et al. (1996). 

\subsection{The Integrated Spectra}
 
As previously said, in addition to the 1D spectra, average 
spectra were obtained for each slit position. For them, the lines 
[O\,{\sc iii}] $\lambda$5007 and [S\,{\sc ii}] $\lambda$6717, as well as H$\beta$ 
and H$\alpha$, were measured. In Table 3, along with the 
line ratios [O\,{\sc iii}]/H$\beta$ and [S\,{\sc ii}]/H$\alpha$, the 
extinction coefficient $C_{\beta}$ and the S/N in H$\alpha$ for the 
integrated spectra are shown. Also, in order to make a quick comparison
with models, the [O\,{\sc iii}]/H$\alpha$ ratio is presented. [N\,{\sc ii}] $\lambda$6583 
was detected in only bh1, while He\,{\sc i} $\lambda$6678 was detected in bh1 and ch1, 
but the values are not presented in the table.

The DIG results obtained from the integrated spectra are very similar to 
those from the 1D spectra: large values of the [O\,{\sc iii}]/H$\beta$ ratio, 
low values of [S\,{\sc ii}]/H$\alpha$, and large extinction. Concerning 
the [O\,{\sc iii}]/H$\beta$ ratio, half of the DIG locations have values 
lower than $1$, but none of them have values lower than $0.5$, which is the 
highest value measured at the DIG in spiral galaxies (Rand 1998). The differences in 
the [O\,{\sc iii}]/H$\beta$ ratio between irregular galaxies and the extraplanar DIG 
in spiral galaxies indicate higher oxygen abundances in irregular galaxies or  
harder spectra. Most of the low excitation values in DDO 53 correspond to the 
southern slits (e-g). On the 
contrary, for the northern slits half of the DIG locations have values of 
[O\,{\sc iii}]/$H\beta$ larger than $1.5$. We have to keep in mind that these 
slits pass through the most luminous H\,{\sc ii} regions of the galaxy, regions No $10$ 
and $13$ (Strobel et al. 1990). Such large values of the [O\,{\sc iii}]/H$\beta$ 
ratio are also measured for the DIG of other irregular galaxies, such as IC 10 (Hidalgo-G\'amez 2005). 
If the absorption-corrected ratios are considered, the [O\,{\sc iii}]/H$\beta$ ratio 
is lower: in general, $50 \%$ lower than the values presented in Table 3. 

We also present the [O\,{\sc iii}]/H$\alpha$ ratio. As previously said, this ratio is given 
because it is widely used in modelling, despite its strong dependence on extinction. 
Therefore, the values presented here should be taken with care. It has been observed that 
this ratio gets larger with the distance from the galactic plane (Otte et al. 2002), with 
values larger than $0.3$ outside the H\,{\sc ii} regions. Values larger than these are 
observed here for almost all DIG locations and H\,{\sc ii} regions. If no extinction correction
is made, the values are even larger. Also, there are no significant changes 
([O\,{\sc iii}]/H$\alpha$ lower then $0.3$) if any 
of the absorption corrections discussed in Section 2 are performed. 
 
There are a few things to be noted in relation to the [S\,{\sc ii}]/H$\alpha$ ratio 
from Table 3. First, the values inside and outside the H\,{\sc ii} regions are very 
similar, as noted for the  1D spectra, and to the contrary of what was found for 
other galaxies (e.g. NGC 6822; Hidalgo-G\'amez \& Peimbert 2007). Second, the values 
increase towards the south (the receding, disturbed part of the galaxy), with a maximum 
at gh2, which is an H\,{\sc ii} region. Eight out of nine integrated spectra 
where sulfur is detected in the southern region have [S\,{\sc ii}]/H$\alpha$ larger than 
$0.1$, as compared to only $3$ out of $10$ in the northern region of DDO 53. The larger 
gradient of the velocity towards the south of the galaxy 
(Begum et al. 2006) might be an indication of a disturbance, which
might create a shock wave. As for higher velocities of the gas the shocks are more important, 
 and this ratio will be larger. These 
results confirm the trend observed in the 1D spectra. An average value of $0.147$ 
could be obtained for the [S\,{\sc ii}]/H$\alpha$ ratio considering all the DIG 
integrated values. This is the second lowest value of this ratio for the sample of 
galaxies already studied, slightly lower than for ESO 245-G05 and GR 8 ($0.165$) and 
much lower than for IC 10 ($0.31$). The low value of the [S\,{\sc ii}]/H$\alpha$ ratio 
might indicate that although the interstellar medium is disturbe d,there is no evidence 
for shocks (see Section 4.1). Again, when the absorption-corrected ratios are used instead, 
the [S\,{\sc ii}]/H$\alpha$ ratio is, on average, $40 \%$ lower than those discussed 
here, but the general trend remains.

Finally, the extinction also follows the same trend that was observed in the 1D 
spectra. DDO 53 has a significant amount of dust. There are locations with $C_{\beta}$ 
larger than $2$, and there is no systematic trend of the H\,{\sc ii} regions being 
dustier than the DIG or vice versa. It is important to note that the southern regions 
seem to be dustier than the northern ones.

\section{The ionization source}

In order to determine the ionization source of the DIG, a comparison can be made between the 
line ratios measured in this galaxy and the values predicted by different models. 
Considering radiation-bounded H\,{\sc ii} regions, Mathis (1986) and Domg\"orgen \& Mathis 
(1994) obtained that the value of both [S\,{\sc ii}]/H$\alpha$ and 
[N\,{\sc ii}]/H$\alpha$ for the DIG are very large, larger than $0.3$, while 
[O\,{\sc iii}]/H$\alpha$ 
is lower than $0.1$. The main caveat is that these results are optimized for Milky Way 
metallicities. As discussed in the next section, the metallicity of DDO 53 is  much 
lower than the Milky Way's. Castellanos et al. (2004) tried to reproduce the line ratios of 
the DIG for low-metallicity galaxies using the photoionization code CLOUDY. They successfully 
reproduced the increase of both [N\,{\sc ii}]/H$\alpha$ and [S\,{\sc ii}]/H$\alpha$, but 
they did not fit the excitation, [O\,{\sc iii}]/H$\beta$. Wood \& Mathis (2004) performed  
Monte Carlo simulations of a multicomponent interstellar medium with different ionizing 
spectra and metallicities in order to explain the increase of the [N\,{\sc ii}]/[S\,{\sc ii}] 
ratio with the distance from the plane. They fit this ratio but not others, such as 
[O\,{\sc iii}]/H$\alpha$. Comparing the value listed in Table 3 with the results from  
Mathis (1986) and Domg\"orgen \& Mathis (1994), it can be concluded that the H\,{\sc ii} 
regions in DDO 53 are not radiation-bounded because [O\,{\sc iii}]/H$\alpha$ is never 
as low as $0.1$, and [S\,{\sc ii}]/H$\alpha$ never as high as $0.3$ except in gh2. 
Actually, the lowest [O\,{\sc iii}]/H$\alpha$ value is $0.20$ (ed2 and ed3).

In recent years density-bounded H\,{\sc ii} regions have been modelled (Wood \& Mathis 2004; 
Hoopes \& Walterbos 2003), but again for high-metallicity regions. Therefore, some corrections 
have to be made. The real corrections cannot be done, but one knows that lower metallicities  
imply a high ionization parameter, which implies higher 
[O\,{\sc iii}]/H$\beta$ ratios. Therefore, the lower the metallicity, 
the higher the [O\,{\sc iii}]/H$\beta$ ratio for similar T$_{ion}$. As a consequence, the 
photon leakage obtained 
from the models might be larger than the real leakage, or T$_{ion}$ lower than the real 
value. 

We compare the values listed in Table 3 with those from the Hoopes \& Walterbos (2003) 
model. This model is preferred because Wood \& Mathis (2004) focus on explaining the 
rising relation between log [N\,{\sc ii}]/H$\alpha$ and log [S\,{\sc ii}]/H$\alpha$ for 
the DIG. [N\,{\sc ii}] is not detected except in a few regions of the DIG in DDO 53, and such 
a relationship cannot be studied. 

Table 4 shows the leakage of photons from the H\,{\sc ii} regions needed for each of the 
line ratios measured for the integrated DIG spectra from the Hoopes \& Walterbos models. 
The corresponding ionization temperatures are also listed. First, we can study the results 
from the [S\,{\sc ii}]/H$\alpha$ ratio. Ionization temperatures lower than $40,000$ K and 
photon leakages lower than $40 \%$ are obtained from them in all the locations where it 
was detected but two (gd3 and fd1). On the contrary, the T$_{ion}$  values which fit 
the [O\,{\sc iii}]/H$\beta$ ratios are always higher than $38,000$ K. Photon leakages 
larger than $40 \%$ are the most common value. Considering that the metallicity of 
DDO 53 is lower than the metallicity of Orion, for which the model is calculated 
(see section 5.1), the values for the [O\,{\sc iii}]/H$\beta$ ratio might be upper 
limits. Therefore, considering the uncertainties of the ratios and the difference
in metallicity, a T$_{ion}$ between $38,000$ and $41,000$ K fits very well both 
ratios for all slit positions. Also, values of photon leakages between $35 \%$ 
and $50 \%$ can account for both ratios in all slits but gd2, where larger leakage is needed. 
Moreover, all slits but a and  b show a good agreement in T$_{ion}$ 
between all the DIG locations inside the slit. The photon leakage range is broader, 
and therefore more difficult to be fitted by a single value.

\begin{figure}
\centering
\includegraphics[width=12cm]{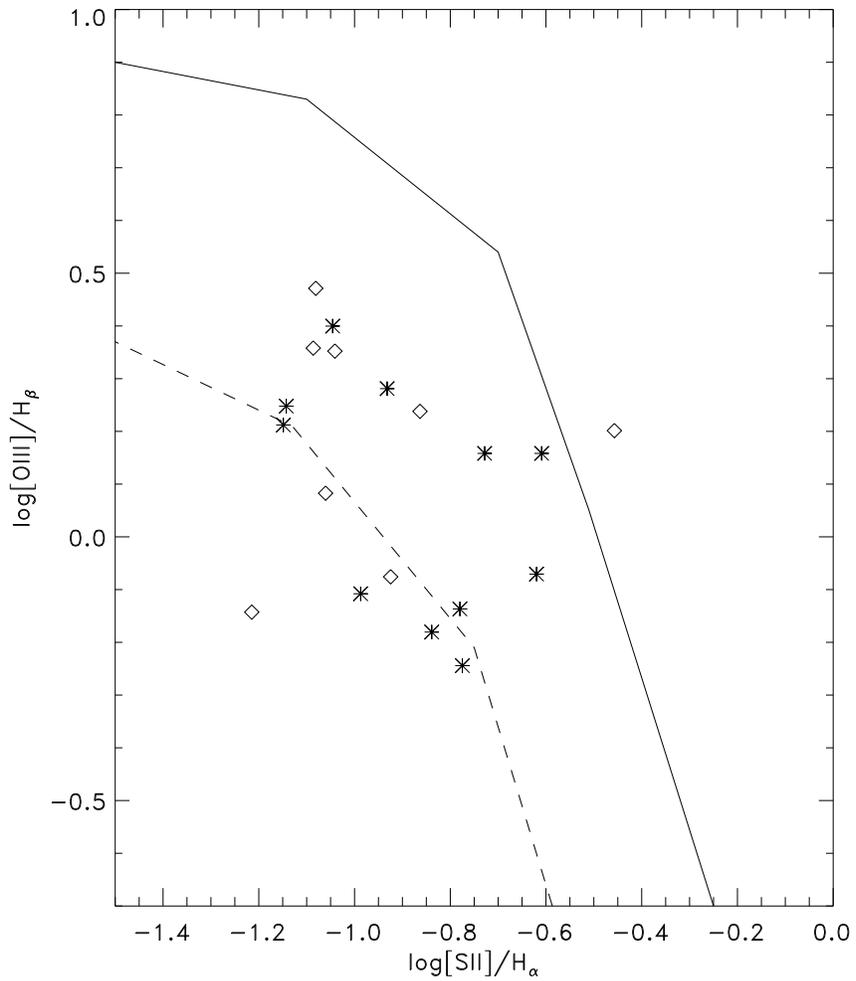}
\caption{[S\,{\sc ii}]/H$\alpha$ vs. [O\,{\sc iii}]/H$\beta$ diagnostic diagram. 
The diamonds correspond to H\,{\sc ii} regions, and asterisk to DIG locations. Superimposed 
are tracks of ionization temperature $50,000$ K ($\it solid line$) and $35,000$ K ($\it dashed line$), 
from photoionization models at low metallicity from Martin (1997).}
\label{fig8}
\end{figure}

These ionization temperatures seem quite realistic, as can be seen from Figure 
~\ref{fig8}, where the  log~[S\,{\sc ii}]/H$\alpha$ versus log~[O\,{\sc iii}]/H$\beta$ 
diagram is shown. Superimposed on the data points (both H\,{\sc ii} regions and 
DIG locations) are tracks of ionization temperature from Figure 3 in Martin (1997). 
The tracks indicate the temperatures needed to ionize the medium only by 
photoionization for a metallicity of $0.20$ (O/H)$_{\odot}$. These are $50,000$ K 
($solid line$) and $35,000$ K ($dashed line$). Most of the DIG data points have 
T$_{ion}$ between $35,000$ and $42,000$ K while one DIG locations have 
T$_{ion}$ higher than $42,000$ K (fd1) and three of them have ionization 
temperatures lower than $35,000$ K (bd2, gd1, and ed2). Location fd1 
has a very large excitation, and therefore, it is located in that position in the 
diagram. Moreover, two H\,{\sc ii} regions (gh2 and eh1) have extreme 
ionization temperatures, larger than $50,000$ K and lower than $35,000$ K, respectively.  
Such values are very unlikely for normal H\,{\sc ii} regions.  Their excitations are 
normal as compared to the values of the other H\,{\sc ii} regions, but they have the 
highest and the lowest [S\,{\sc ii}]/H$\alpha$ ratios (see Table 3), and therefore, they 
occupy these positions in the log~[S\,{\sc ii}]/H$\alpha$ versus log~[O\,{\sc iii}]/H$\beta$ 
diagram. From Figure ~\ref{fig8} it is also interesting to note that there is no 
real difference in the ionization temperatures between H\,{\sc ii} regions and DIG locations.   

It can be concluded that most of the locations agree with T$_{ion}$ values between 
$35,000$ and $41,000$ K and photon leakages between $35 \%$ and $50 \%$. The uniformity 
of these values is very interesting, indicating similar conditions throughout the 
galaxy. There cannot be a single value for the whole galaxy, as physical conditions 
vary throughout it. Moreover, they are independent of absorption correction, in the 
sense that 
the same values are obtained for the T$_{ion}$ and leakage percentages when the 
absorption-corrected ratios are considered. With these values, and following Vacca et al. (1996), the  
stars responsible for the ionization are of types O7-O9, with masses between 20 and 
30 $M_{\odot}$.

\subsection{Other Sources of Ionization?}

As we said before, photon leakages from H\,{\sc ii} regions seem to be the main ionization 
source of the DIG in DDO 53. In any case, it might be interesting to look for other 
ionization sources. 

\begin{figure}
\centering
\includegraphics[width=12cm]{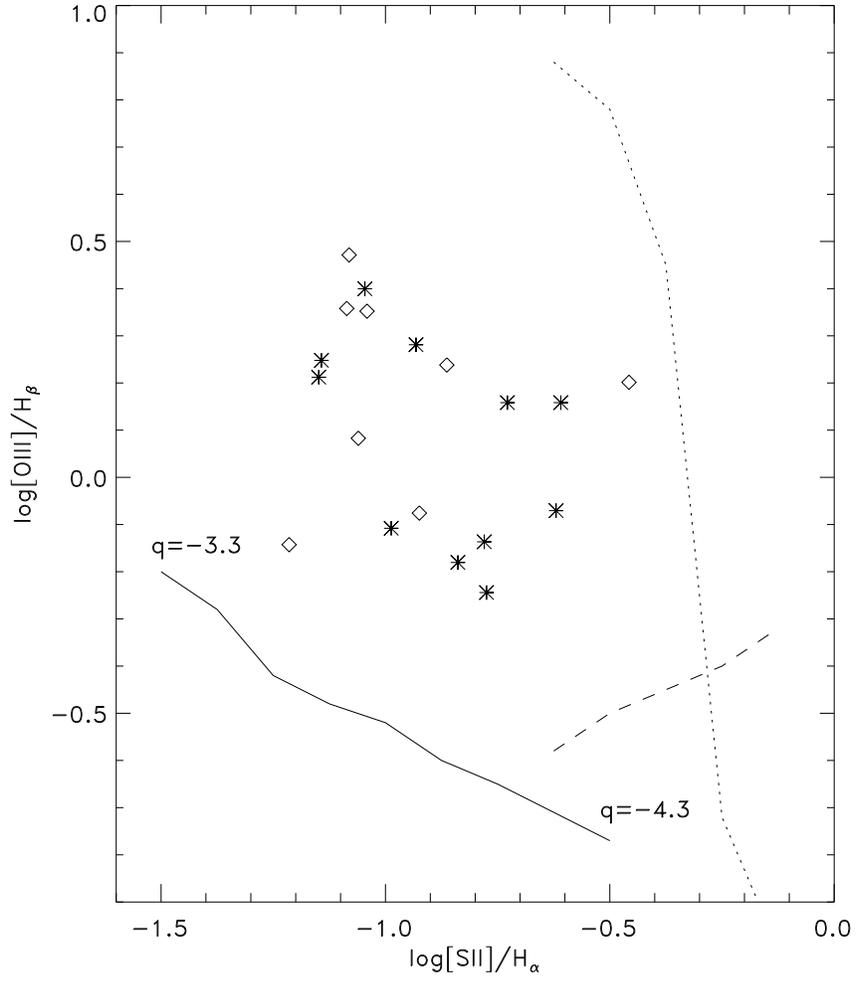}
\caption{[S\,{\sc ii}]/H$\alpha$ vs. [O\,{\sc iii}]/H$\beta$ diagnostic diagram. The diamonds 
correspond to H\,{\sc ii} regions, and asterisk to DIG locations. Superimposed are tracks of different
 models: density-bounded ($\it solid line$), radiation-bounded ($\it dotted line$) and turbulent mixing layers 
($\it dashed line$) from Rand (1998).  }
\label{fig9}
\end{figure}

Turbulent mixing layers (TMLs) are good candidates in the sense that the low-ionization 
lines, such as [S\,{\sc ii}] $\lambda$6717, are the most affected by such mechanisms. Shear flows at 
the boundaries of hot and cold gas produce intermediate-temperature gas (T $\approx$ 10$^5$) 
that radiates strongly in the optical, ultraviolet, and extreme-ultraviolet. By including the 
effects of non-equilibrium ionization and self-photoionization of the gas as it cools after 
mixing, the intensity of some lines in the optical, infrared, and ultraviolet can be predicted. 
In Figure ~\ref{fig9} we have plotted the data points for DDO 53 along with the TML model by 
Slavin et al. (1993) as a dashed line, the radiation-bounded model ($dotted line$), and the 
density-bounded model ($\it solid line$) from Rand (1998). The dispersion is large, especially 
when H\,{\sc ii} regions are considered, but it can be concluded that TML is not playing a 
part in DDO 53, as expected. It is interesting to see that all the data points are between 
the density-bounded and the radiation-bounded lines, as expected. The slope of the DIG data 
points (-$0.395$) is closer to the slope of the density-bounded model (-$0.439$). On the contrary, the 
slope of the H\,{\sc ii} regions is very different ($0.257$). In addition, it can be seen that 
most of the data points agree with an ionization factor of $q \le -3.5$, in agreement with the 
value we used in the Hoopes \& Walterbos model of $-3$ [$q$ is related to the ionization
parameter as $U = 0.0013 ({T/10^4})^{-0.55} q^{1/3}$ according 
to Hoopes \& Walterbos (2003)]. 

\begin{figure}
\centering
\includegraphics[width=12cm]{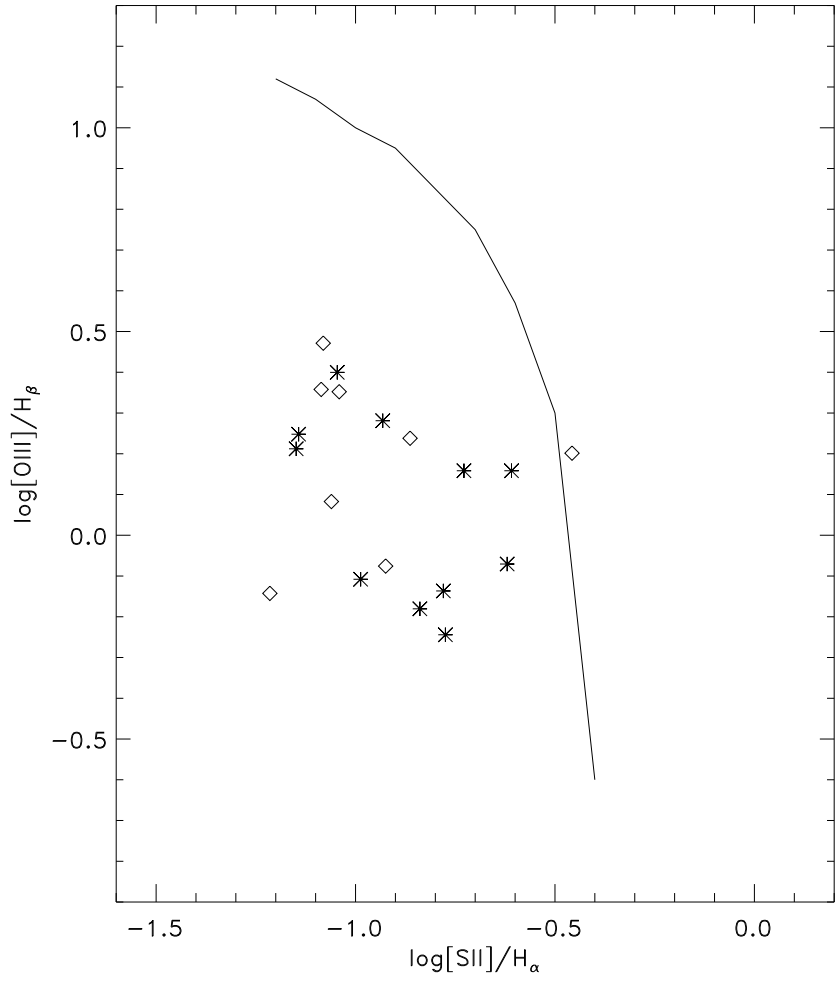}
\caption{[S\,{\sc ii}]/H$\alpha$ vs. [O\,{\sc iii}]/H$\beta$ diagnostic diagram. 
The diamonds correspond to H\,{\sc ii} regions, and asterisk to DIG locations. The solid 
line divides the diagram into the shocked regions (to the left) and the photoionization 
regions (to the right).}
\label{fig10}
\end{figure}

Finally, it might be interesting to explore the existence of shock waves, 
which might have some influence on the ionization. As previously said, the interstellar
medium  of 
this galaxy seems to be quite disturbed, with a gradient in the velocity increasing 
towards the south. Such disturbances might create shock waves which can be amplified 
along the galaxy. As we previously discussed, there is only one location where the  
[S\,{\sc ii}]/H$\alpha$ ratio is  larger than $0.3$ (7h1). In any case, the best 
way to check the existence of shocks is the diagnostic diagram log[S\,{\sc ii}]/H$\alpha$ 
versus log[O\,{\sc iii}]/H$\beta$ (Veilleux \& Osterbrock 1987). Figure ~\ref{fig10} shows 
both the DIG and H\,{\sc ii} region data points. To the left of the solid line there is 
the photoionized region, and to the right the shocked region. Only one H\,{\sc ii} region 
(gh2) is in this latter part of the diagram. The rest of the locations are in the 
photoionization region. Therefore, it can be concluded that shocks are not important in 
this galaxy despite the disturbances of the interstellar medium.

\section{Discussion}

From the discussion in the previous section it is obvious that leakage photons from the 
H\,{\sc ii} regions are the main cause of the ionization of the DIG in DDO 53. The 
values of the leakage are between $35\%$ and $50\%$, which are similar to those in other 
dwarf irregular galaxies. No other ionization sources (shock waves, mixing layers, 
etc) seem to be at work in this galaxy. 

We can compare the line ratios obtained in DDO 53 with those from other galaxies  
studied so far. The DIG [S\,{\sc ii}]/H$\alpha$ ratio is very similar in all the galaxies, 
with values 
of $0.12$-$0.16$ except IC 10 ($0.31$; Hidalgo-G\'amez 2005). On the contrary, the DIG 
excitation changes a lot from galaxy to galaxy. The larger value corresponds to IC 10 
($1.66$), but both NGC 6822 and DDO 53 have values very similar to this one. The lowest 
excitation corresponds to ESO 245-G05 ($0.81$; Hidalgo-G\'amez 2006). These high values 
of the excitation can be due to differences in the oxygen content or in the ionizing 
spectra. DDO 53 has a very similar oxygen content to ESO 245-G05 of 12+log(O/H) = $7.8$,  
but their [O\,{\sc iii}]/H$\beta$ ratios are very different. As we said, the value in 
DDO 53 is similar to those in NGC 6822 and IC 10, which are much more metallic. Therefore, 
differences in the metallicity cannot account for the differences in excitation. A harder spectrum 
might indicate more massive, hotter stars. As a consequence, the  H\,{\sc ii} regions might 
be larger and more luminous. Strobel et al. (1990) concluded that the value of the slope of 
the luminosity function in H$\alpha$ is similar to those found in other irregular galaxies, 
such as NGC 6822, IC 10, and GR 8. They reached a similar conclusion on the sizes of the 
H\,{\sc ii} regions. These four galaxies have very similar values of the DIG 
[O\,{\sc iii}]/H$\beta$ ratio, indicating that their ionizing spectra are very similar. 
Therefore, it can be concluded that the excitation values at the DIG are more closely 
related to the ionizing spectra than to the oxygen content.

\subsection{The Metallicity of DDO 53}

In addition to the study of the DIG we can determine the chemical abundances in DDO 53 from the
intensities of the integrated H\,{\sc ii} spectra. The main caveat 
is the low efficiency of the Boller\&Chives spectrograph below $4000$ \AA. 
As mentioned previously, it is less 
than $10\%$. Therefore, although the intensities of the lines below $4000$ \AA~ could 
be detected, the values 
might not be very realistic. For example, the corrected intensity of the 
[O\,{\sc ii}]/H$\beta$ ratio of region ch1 is 
0.20, which is very low when compared with the values in other H\,{\sc ii} regions 
in dwarf irregular 
galaxies (see Hidalgo-G\'amez \& Olofsson 2002). Therefore, in order to obtain more 
reliable intensities 
of the [O\,{\sc ii}] lines, equation (1) in Hidalgo-G\'amez \& Ramirez-Fuentes (2007) is used. This equation 
relates  the intensities of [O\,{\sc ii}]$\lambda$3727 and [O\,{\sc iii}]$\lambda$5007. The 
reliability and uncertainties of this equation are discussed in Hidalgo-G\'amez \& Ramirez-Fuentes (2007). 
In this case, we feel that the values obtained from this equation are more trustworthy than those measured 
in the spectra, and these will be the values used in the determination of the chemical abundances presented 
here.

In two of these H\,{\sc ii} integrated spectra, the forbidden oxygen line 
[O\,{\sc iii}] $\lambda$4363 is 
present, so we can obtain a reliable determination of the electron temperature and the chemical 
abundances following the so-called standard method (Osterbrock 1989). These two spectra are  
regions $2h1$ and $3h1$ in Table 2, which correspond to regions No $10$ and No $13$ in Strobel et al. 
(1990). The intensities of all the lines measured, extinction-and absorption-corrected, are given in 
Table 5, along with their uncertainties. 

The determined oxygen and nitrogen abundances for regions No 10 and No 13 are presented in 
Table 6, along with the electronic temperature of the O$ ^{++}$ (inner) and the N$ ^+$ (outer) zones. The
oxygen abundances obtained here of $7.80\pm0.1$ and $7.91\pm0.3$ agree with the values 
of =$7.8$ obtained 
with semiempirical methods (I. Saviane 2004, private communication), but they 
are slightly  higher than the previous values obtained for this galaxy of $7.62$ (Skillman et al. 1989). 
In any case, DDO 53 is a low-metallicity galaxy. It is confirmed by the low nitrogen 
abundance of 
12+log(N/H) = $5.7$. Moreover, nitrogen is not detected in region No 13. None of the 
previous determinations
of the abundances of DDO 53 measured nitrogen; therefore, a comparison cannot be made. This value is 
even lower than for DDO 190 [12+log(N/H) = $5.97$; Hidalgo-G\'amez  \& Olofsson 2002] which is the 
galaxy with the lowest nitrogen abundance in a sample of dwarf irregular galaxies. Moreover, the 
log(N/O) ratio is also very low, only $-2.12$, which is very much lower than the canonical $-1.5$ 
obtained for a sample of blue compact and irregular galaxies (Izotov \& Thuan 1999). These low 
nitrogen abundances might explain the lack of this line in the majority of the H\,{\sc ii} regions and 
DIG locations. Such low values of nitrogen have been observed in other dwarf irregular galaxies, such as 
GR 8 [log(N/O) = -1.77; Moles et al. 1990].  

Finally, we can also determine the helium abundance. Only one line is used (He\,{\sc i} at  6678 \AA),
because the intensities of other lines can be polluted by sky lines (He\,{\sc i} at 5578 \AA) or by 
processes other than pure recombination (He\,{\sc i} at 4471 \AA; Izotov \& Thuan 2004). Moreover, 
as the most reliable ionization correction factor is based on the intensity of [S\,{\sc iii}] lines,
which are not present in our spectra, the He$^+$/H$^+$ ratio is given.  The values of both regions 
No 10 and No 13
are equal to $0.11$, very similar to that of other galaxies of the same type (Hidalgo-G\'amez 1999). 

The abundances of these three elements follow a similar pattern to those of DDO 190, with very low nitrogen
and low oxygen but normal helium values (see Table 3 in Hidalgo-G\'amez \& Olofsson 2002). 

Another interesting problem is the (in)homogeneity of the chemical abundances in dwarf irregular 
galaxies. Only in three galaxies have enhancements of oxygen (IC 4662 and ESO 245-G05; Hidalgo-G\'amez et 
al. 2001) or nitrogen (NGC 5253; Kobulnicky et al. 1997) been reported. We might use the values 
presented here to study the oxygen distribution along DDO 53. Regions No 10 and No 13 are quite 
close to each other, and their oxygen and helium abundances are similar considering the uncertainties. In other
to enlarge the number of oxygen determinations we can use the other seven H\,{\sc ii} regions defined 
in Section 3.1. Again, the [O\,{\sc ii}]$\lambda$3727 intensity can be obtained from equation (1) in  
Hidalgo-G\'amez \& Ram\'{\i}rez-Fuentes (2007). With the intensities, reddening-corrected and normalized 
to H$\beta$, of [O\,{\sc ii}]$\lambda$3727, [O\,{\sc iii}]$\lambda$4959, and 
[O\,{\sc iii}]$\lambda$5007, any of the semiempirical methods, the $R_{23}$ (Pagel et al. 1979) 
or the $P$ method (Pilyugin 2000, 2001), can be used to determine the oxygen abundance. The values are shown 
in Table 7 along with the uncertainties. These uncertainties take into account the intrinsic 
uncertainties of the methods themselves due to their statistical nature. Such uncertainties might be 
$0.2$ dex according to Hidalgo-G\'amez \&  Ram\'{\i}rez-Fuentes (2007). Considering all the uncertainties, 
the oxygen abundances reported in Table 7 should be considered as a first approximation only. In 
consequence, few conclusions can be obtained. First, DDO 53 seems to be a low metallicity galaxy. 
But at the center of the galaxy, the metallicity seems to be even lower (regions No 5,  No 6, and 
No 11) than in the rest of the H\,{\sc ii} regions, the difference being $0.6$ dex, much larger 
than the uncertainties. In order to clarify this point, new measurements with good resolution and 
high S/N ranging from [O\,{\sc ii}] $\lambda$3727 to [S\,{\sc ii}] $\lambda\lambda$6717, 6731 are needed 
in order to get a better determination of the abundances throughout the galaxy. Such observations 
would throw light on this situation.  

\section{Conclusions}

We studied the line ratios of the DIG in the dwarf irregular galaxy DDO 53 using long-slit spectroscopy. 
The DIG values were compared with those of the H\,{\sc ii} regions. In order to do this comparison,  
definitions of H\,{\sc ii} regions and the DIG were made based on the flux of the H$\alpha$ line. With this 
definition we divided the data into $9$ H\,{\sc ii} regions and 17 DIG locations. The differences in 
the line ratios among them are less extreme than for other dwarf irregular galaxies studied before, such as 
NGC 6822 (Hidalgo-G\'amez \& Peimbert 2007) and IC 10 (Hidalgo-G\'amez 2005). The excitation is as 
large in this galaxy, especially at the DIG location, as in other galaxies, indicating similar ionizing 
spectra. On the contrary, the [S\,{\sc ii}]/H$\alpha$ ratio is lower than in the previously studied 
galaxies. [N\,{\sc ii}] $\lambda$6583 is detected in very few locations along the slits and is not 
studied here. 
In addition to the line ratios, the extinction is studied along the slit positions, indicating that the 
dust inside the northern part of the galaxy is more homogeneously distributed than in the southen part, 
where large clumps of dust are clearly detected. 

The line ratios obtained are consistent with density-bounded H\,{\sc ii} regions with a photon leakage 
of $35 \%$ -$50 \%$ and ionization temperatures between  $35,000$ and $42,000$ K for most of the 
locations. No other sources of ionization (shocks, turbulence, etc,) are needed in order to explain the 
line ratios observed in spite of the disturbed kinematics of the galaxy in H\,{\sc i} (Begum et al. 
2006). 

Finally, we determined the abundances for two H\,{\sc ii} regions using the electron temperature method. 
The values are very similar and indicate that this is a low-metallicity galaxy. When semiempirical 
methods are used to determine the oxygen abundances for a large number of H\,{\sc ii} regions, a difference 
towards the center of the galaxy is obtained, indicating possible inhomogeneities in the oxygen content. 
In any case, new spectra are needed in order to confirm this trend.

The author is indebted to Nahiley Flores-Fajardo for preparing Figure 8. She thanks 
G. Garc\'{\i}a-Segura for interesting discussions. The anonymous referee is thanked for interesting 
comments and discussion which have improved this manuscript. She also thanks J. Brenan for a 
careful reading of the manuscript. This investigation is supported by DGAPA project IN114107.

\clearpage

\clearpage

\clearpage

\clearpage

\clearpage

\clearpage

\clearpage

\clearpage

\clearpage

\clearpage

\begin{table}
\caption[]{Log of the Observations. 
The slit positions observed are presented in column 1. 
Column 2 shows the date of observation while the seeing is given in 
column 3 in arcsecond. The total integration time in minutes is presented in 
column 4 and the air mass in column 5. Column 6 gives the telescope 
coordinates for the first slit position for each night while the rest of the positions 
is given by the number of arcseconds 
the telescope was moved from the initial position. }
\vspace{0.05cm}
\begin{center}
\begin{tabular}{c c c c c c}
\hline
{Position}  & {Date}  &
{Seeing} & {Int. Time} & {Air Mas. } & {Comments} \\ 
\hline 
slit a        & 12032002 & 1.5'' & 60$^m$   & 1.3 & 08$^h$34$^m$ 66$^o$10'\\
slit b        & 12032002 & 1.5'' & 85$^m$   & 1.2 & 7$^{''}$ towards the south\\
slit c        & 13032002 & 1.5'' & 130 $^m$ & 1.2 & 08$^h$34$^m$18$^s$ 66$^o$09`51``\\
slit d        & 13032002 & 1.5'' & 15 $^m$  & 1.2 &  6'' towards the south\\
slit e        & 14032002 & 2''   & 70 $^m$  & 1.2 & 08$^h$34$^m$15$^s$ 66$^o$09`25`` \\
slit f        & 14032002 & 2''   & 30 $^m$  & 1.2 &  6'' towards the south\\
slit g        & 14032002 & 2''   & 30 $^m$  & 1.2 &  12'' towards the south \\
slit h        & 14032002 & 2''   & 30 $^m$  & 1.3 &  18'' towards the south\\
\hline
\end{tabular}
\end{center}
\end{table}

\clearpage

\begin{table}
\caption[]{Total uncertainties in the lines ratios studied for each slit position for 
the DIG locations (at the top) and the H\,{\sc ii} regions. Three different terms have 
been considered: the uncertainty in the continuum, the uncertainty due to the reduction 
and calibration procedure, and the uncertainty due to the extinction correction. For the 
blended lines an extra term has been considered. Those line ratios not detected at one 
slit location are marked with -.} 
\vspace{0.05cm}
\begin{center}
\begin{tabular}{ c c c c c}
\hline
{Slit} & {[O\,{\sc iii}]/H$\beta$} & [O\,{\sc iii}]/H$\alpha$ & {[N\,{\sc ii}]/H$\alpha$} & {[S\,{\sc ii}]/H$\alpha$}  \\
\hline
slit a & 4\% & 3\%  & 12\% & - \\
slit b & 3\% & 4\%  & 12\% & 12\% \\
slit c & 2\% & 2\% & 12\% & 12\% \\
slit d & 2\% & 2\% &  -   & 12\% \\
slit e & 3\% & 3\% & 24\% & 13\% \\
slit f & 5\% & 5\% & 12\% & 13\% \\
slit g & 4\% & 3\% & 12\% & 14\% \\
slit h & 5\% & 5\% & 15\% & 16\% \\
\hline
slit a & 2\% & 2\% & 10\% & - \\
slit b & 2\% & 2\% & 10\% & 10\% \\
slit c & 2\% & 2\% & 17\% & 10\% \\
slit d & 2\% & 2\% & 14\% & 10\% \\
slit e & 2\% & 2\% & 14\% & 11 \% \\
slit f & -   &  -  &  -   & -\\
slit g & 2\% & 2\% & 10\% & 10\% \\
slit h & 2\% & 2\% & 10\% &  - \\    
\hline
\end{tabular}
\end{center}
\end{table}

\clearpage

\begin{table}
\caption[]{Extinction-corrected line ratios of the integrated spectra for DDO 53. Column 1 shows 
the location, where the 
first character is the slits position from Table 1 and the second is ``h'' for H\,{\sc ii} 
region and ``d'' for DIG location. The third character is 1 for the first (or only) location 
in the slit and higher for additional locations. Columns 2, 3 and 4 show the [O\,{\sc iii}]/H$\beta$
ratio, the [O\,{\sc iii}]/H$\alpha$ ratio and  the [S\,{\sc ii}]/H$\alpha$ ratio when detected. 
Finally, column 5 shows the extinction coefficient in the recombination line H$\alpha$ and 
the S/N ratio in H$\alpha$ is given in column 6. }
\vspace{0.05cm}
\begin{center}
\begin{tabular}{c c c c c c}
\hline
{Location} & {[O\,{\sc iii}]/H$\beta$} & { [O\,{\sc iii}]/H$\alpha$} & {[S\,{\sc ii}]/H$\alpha$} & { C$\beta$} & {S/N(H$\alpha$)}\\
\hline240
ad1   &  1.77$\pm$0.07   &  0.62$\pm$0.03    & 0.072$\pm$0.01   &  2.00$\pm$0.06  &  21 \\
ah1   &  1.21$\pm$0.05   &  0.42$\pm$0.03    & 0.087$\pm$0.007  &  1.72$\pm$0.07  &  61 \\
ad2   &  1.35$\pm$0.2    &      -             &      -           &  -             &   3 \\
bd1   &  1.91$\pm$0.06   &  0.67$\pm$0.03    & 0.117$\pm$0.007  &  0.99$\pm$0.08  &  18 \\
bh1   &  2.96$\pm$0.05   &  1.03$\pm$0.02    & 0.083$\pm$0.007  &  1.58$\pm$0.06  & 188 \\
bd2   &  0.78$\pm$0.09   &  0.27$\pm$0.03    & 0.103$\pm$0.02   &  1.07$\pm$0.07  &  10 \\
cd1   &  1.63$\pm$0.2    &       -           & 0.071$\pm$0.01   &   -             &  13 \\
ch1   &  2.28$\pm$0.09   &  0.80$\pm$0.03    & 0.082$\pm$0.007  &  1.59$\pm$0.06  &  55 \\
cd2   &  0.73$\pm$0.1    &  0.25$\pm$0.04    & 0.166$\pm$0.03   &  5.54$\pm$0.8   &   4 \\
dd1   &  2.51$\pm$0.08   &  0.88$\pm$0.03    & 0.090$\pm$0.007  &  0.44$\pm$0.03  &  18 \\
dh1   &  2.25$\pm$0.08   &  0.79$\pm$0.03    & 0.091$\pm$0.006  &  1.58$\pm$0.06  &  81 \\
dd2   &  1.79$\pm$0.2    &  0.63$\pm$0.07    &        -         &  1.92$\pm$0.1   &   9 \\
ed1   &  1.44$\pm$0.2    &  0.50$\pm$0.07    & 0.187$\pm$0.03   &  1.86$\pm$0.3   &   3 \\
eh1   &  0.72$\pm$0.02   &        -          & 0.061$\pm$0.005  &        -        &  20 \\
ed2   &  0.57$\pm$0.06   &  0.20$\pm$0.02    & 0.168$\pm$0.02   &  2.55$\pm$0.3   &  8 \\
eh2   &  0.84$\pm$0.03   &  0.29$\pm$0.01    & 0.119$\pm$0.02   &  2.93$\pm$0.1   &  26\\
ed3   &  0.56$\pm$0.07   &  0.20$\pm$0.02    &      -           &  2.45$\pm$0.3   &  9\\
fd1   &  1.44$\pm$0.2    &  0.51$\pm$0.07    & 0.246$\pm$0.04   &  1.96$\pm$0.3   &  4\\
gd1   &  0.66$\pm$0.08   &  0.23$\pm$0.03    & 0.145$\pm$0.03   &  2.29$\pm$0.1   & 10\\
gh1   &  1.73$\pm$0.06   &  0.61$\pm$0.02    & 0.137$\pm$0.02   &  1.54$\pm$0.03  & 29\\
gd2   &  0.94$\pm$0.1    &  0.33$\pm$0.04    &  -               &  2.35$\pm$0.3   & 13\\
gh2   &  1.59$\pm$0.2    &       -           & 0.349$\pm$0.05   &   -             & 15\\ 
gd3   &  0.85$\pm$0.1    &  0.30$\pm$0.04    & 0.24$\pm$0.05    &  0.84$\pm$0.04  & 10\\
hd1   &  1.55$\pm$0.2    &  0.54$\pm$0.08    &    -             &  2.71$\pm$0.4   &  3\\
hh1   &  3.79$\pm$0.1    &       -           &   -              &      -          & 31 \\
hd2   &  1.26$\pm$0.2    &       -           &   -              &        -        &  3\\
\hline   
\end{tabular}
\end{center}
\end{table}

\clearpage

\begin{table}
\caption[]{Photon leakage (in $\%$) and ionization temperatures for the  
integrated spectra for DDO 53 based on the line ratios. The locations are 
named as in Table 3. The second and third columns correspond to the data 
from the excitation and the fourth and fifth are based on the 
[S\,{\sc ii}]/H$\alpha$ ratio, when detected.   }
\vspace{0.05cm}
\begin{center}
\begin{tabular}{c c c c c}
\hline
{Location}  & {T$_{ion}$  K} & {Leakage [O\,{\sc iii}] ($\%$)} & {T$_{ion}$  K} & {Leakage [S\,{\sc ii}]($\%$) }   \\ 
\hline 
ad1  & 38,000-50,000 & 30-90 $\%$ & 30,000-34,000 & 30-35 $\%$ \\
ad2  & 38,000-41,000 & 40-70 $\%$ & -             &  - $\%$  \\
bd1  & 41,000-50,000 & 50-80 $\%$ & 30,000-36,000 & 30-38 $\%$ \\
bd2  & 38,000        & 52    $\%$ & 30,000-36,000 & 30-40 $\%$  \\
cd1  & 38,000-47,000 & 35-80 $\%$ & 30,000-34,000 & 30-35 $\%$ \\
cd2  & 38,000        &  60   $\%$ & 30,000-38,000 & 30-42 $\%$ \\
dd1  & 41,000-50,000 & 42-65 $\%$ & 30,000-34,000 & 30-35 $\%$  \\
dd2  & 41,000-50,000 & 50-90 $\%$ &  -            &   -   \\
ed1  & 38,000-44,000 & 38-90 $\%$ & 30,000-40,000 & 30-45 $\%$ \\
ed2  & 38,000        & 72 $\%$    & 30,000-38,000 & 30-42 $\%$  \\
ed3  & 38,000        & 75 $\%$    & -             &  -    \\
fd1  & 38,000-41,000 & 38-85 $\%$ & 30,000-40,000 & 35-50/70-80 $\%$ \\
gd1  & 38,000        & 62 $\%$    & 30,000-38,000 & 30-42 $\%$    \\
gd2  & 38,000-41,000 & 50-90 $\%$ &    -          & -     \\
gd3  & 38,000        & 55  $\%$   & 30,000-40,000 & 35-50/70-80 $\%$  \\
hd1  & 38,000-44,000 & 35-85 $\%$ &   -           &   -            \\
hd2  & 38,000-41,000 & 42-68 $\%$ &     -         &    -         \\  
\hline
\end{tabular}
\end{center}
\end{table}

\clearpage

\begin{table}
\caption[]{Line intensities, normalized to H$\beta$, detected inside two of the 
H\,{\sc ii} regions in DDO 53. The values are absorption and extinction corrected. }
\vspace{0.05cm}
\begin{center}
\begin{tabular}{c c c}
\hline
{Line} & {region No 10} & {region No 13}\\
\hline
$\lbrack$O\,{\sc ii}$\rbrack\lambda$3727      & 4.6$\pm$0.5   & 0.21$\pm$0.01 \\
$\lbrack$Ne\,{\sc iii}$\rbrack$ $\lambda$3970 & 0.20$\pm$0.01 & 0.048$\pm$0.001 \\
H$\delta$ $\lambda$4102               & 0.20$\pm$0.01 & 0.142$\pm$0.003 \\
H$\gamma$ $\lambda$4340               & 0.47$\pm$0.01 & 0.40$\pm$0.01 \\
$\lbrack$O\,{\sc iii}$\rbrack$$\lambda$4363   & 0.056$\pm$0.002 & 0.033$\pm$0.002 \\
He\,{\sc i} $\lambda$4471                     & 0.043$\pm$0.001 & -   \\
H$\beta$ $\lambda$4865                & 1.00$\pm$0.02 & 1.00$\pm$0.02 \\
$\lbrack$O\,{\sc iii}$\rbrack$$\lambda$4959   & 0.96$\pm$0.04 & 0.78$\pm$0.02  \\
$\lbrack$O\,{\sc iii}$\rbrack$$\lambda$5007   & 2.79$\pm$0.09 & 2.27$\pm$0.05\\
$\lbrack$N\,{\sc i}$\rbrack$$\lambda$5198      & 0.021$\pm$0.001 & 0.044$\pm$0.001\\
He\,{\sc i} $\lambda$5875                     & 0.08$\pm$0.01  & 0.01$\pm$0.007\\
H$\alpha$ $\lambda$ 6568              & 2.86$\pm$0.06  & 2.86$\pm$0.06 \\
$\lbrack$N\,{\sc ii}$\rbrack$$\lambda$6583    & 0.026$\pm$0.004 & -  \\
HeI $\lambda$6678                     & 0.029$\pm$0.001 & 0.030$\pm$0.01\\
$\lbrack$S\,{\sc ii}$\rbrack$$\lambda$6617    & 0.136$\pm$0.003 & 0.165$\pm$0.004 \\
$\lbrack$S\,{\sc ii}$\rbrack$$\lambda$6631    & 0.100$\pm$0.002 & 0.119$\pm$0.003\\
\hline
\end{tabular}
\end{center}
\end{table}

\clearpage

\begin{table}
\caption[]{Electron temperatures, Oxygen and Nitrogen abundances as well as Helium abundances 
of two of the H\,{\sc ii} regions in DDO 53, determined with the standard method. }
\vspace{0.05cm}
\begin{center}
\begin{tabular}{c c c}
\hline
{Line} & {region No 10} & {region No 13}\\
\hline
T$_e$ (O$^{++}$) & 15214 & 13258\\
T$_e$ (O$^+$)  & 13650 & 12281\\
12+log(O/H) & 7.8$\pm$0.1 & 7.91$\pm$0.3\\
12+log(N/H) & 5.67$\pm$0.2 &   - \\
log(N/O)    & -2.12$\pm$0.2 &   - \\
He$^+$/H$^+$      & 0.11$\pm$0.01   & 0.11$\pm$0.03 \\
\hline
\end{tabular}
\end{center}
\end{table}

\clearpage

\begin{table}
\caption[]{Oxygen Abundances of the H\,{\sc ii} regions in DDO 53. The identification of the H\,{\sc ii} region according to Strobel et al. (1990) is given in column 1. The oxygen abundances based on the standard method with the electronic temperature are given in column 2 for the only two regions where the [O\,{\sc iii}]$\lambda$4363\AA~ line was detected. The oxygen abundances determined with the R$_{23}$ calibrator by Pagel et al. (1979) are shown in column 3 while the abundances determined with the $P$ calibrator by Pilyugin (2001) are shown in column 4.  } 
\vspace{0.05cm}
\begin{center}
\begin{tabular}{c c c c}
\hline
{H\,{\sc ii} region} & {12+log(O/H)} & {R$_{23}$} & {P$_b$}\\
\hline
No 9 &    -           &   7.53$\pm$0.4 & 7.48$\pm$0.3\\
No 10 & 7.80$\pm$0.1  & 8.07$\pm$0.3 & 7.95$\pm$0.2\\
No 13 & 7.91$\pm$0.3  & 7.98$\pm$0.5 & 7.94$\pm$0.4\\
No 5  &   -           &   7.33$\pm$0.4 & 7.23$\pm$0.3\\
No 6  &   -           &   7.01$\pm$0.5 & 6.76$\pm$0.4\\
No 11 &   -           &   7.17$\pm$0.4 & 6.98$\pm$0.3\\
No 8  &   -           &   7.82$\pm$0.4 & 7.82$\pm$0.4\\
No 16 &   -           &   7.76$\pm$0.5 & 7.76$\pm$0.4\\
No 17 &   -           &   8.10$\pm$0.4 & - \\
\hline
\end{tabular}
\end{center}
\end{table} 


\begin{thebibliography}{}

\bibitem[2006]{begum} Begum, A., Chengalur, J.N., Karachentsev, I.D., Kaisin, S.S., \& Sharina, M.E. 2006, MNRAS, 365, 1220

\bibitem[1976]{benvenuti} Benvenuti, P., D'Odorico, S., \& Peimbert, M. 1976, RevMexA\&A, 2, 3

\bibitem[1971]{brocklehurst} Brocklehurst, M. 1971, MNRAS, 153, 471

\bibitem[1996]{caplan} Caplan, J., Ye, T., Deharveng, L., Turtle, A.J., \& Kennicutt, R.C. 1996, A\&A, 307, 403 


\bibitem[2003]{castellanos} Castellanos, M., Valls-Gabaud, D., D\'{\i}az, A. \& Tenorio-Tagle, G. 2004, ''How does the Galaxy work ? `` eds: E.J. Alfaro, E. P\'erez \& J. Franco, Kluwer Academic Publishers, p. 101

\bibitem[1994]{domgonge} D\"omgonge, H. \& Mathis. J.S. 1994, ApJ, 428, 647

\bibitem[1993]{dopitaa} Dopita, M. A. 1993, PASAu, 10, 359

\bibitem[1998]{geenawalt} Greenawalt, B., Walterbos, R.A.M., Thilker, D. \& Hoopes, C.G. 1998, ApJ 506, 135


\bibitem[1999]{Hidalgo-Gameza} Hidalgo-G\'amez, A.M. 1999, Ph.D. Thesis, Uppsala University

\bibitem[2005]{Hidalgo-Gamezb} Hidalgo-G\'amez, A.M., 2005, A\&A 442, 443

\bibitem[2005]{hidalgo-Gamezc} Hidalgo-G\'amez, A.M. 2006, AJ, 131, 2078

\bibitem[1998]{hidalgo-Gamezd} Hidalgo-G\'amez, A.M., \& Olofsson, K. 1998, A\&A, 334, 45

\bibitem[2001]{Hidalgo-Gameze} Hidalgo-G\'amez, A.M., Masegosa, J., \& Olofsson, K. 2001, A\&A, 369, 797

\bibitem[2002]{Hidalgo-Gamezf} Hidalgo-G\'amez, A.M., \& Olofsson, K. 2002, A\&A, 389, 836

\bibitem[2006]{hidalgo-Gamezg} Hidalgo-G\'amez, A.M. \& Peimbert, A. 2007, AJ, 133, 1874
 
\bibitem[2003]{hoopes} Hoppes, C. G. \& Walterbos, R.A.M. 2003, ApJ, 586, 902

\bibitem[1992]{hunter} Hunter, D.A. \& Gallagher, J.S. III 1992, ApJ, 391, 9

\bibitem[2004]{huntera} Hunter, D.A. \& Elemegreen, B.G. 2004, AJ, 128, 2170

\bibitem[1999]{izotov} Izotov, Y.I. \& Thuan, T.X. 1999, ApJ, 511, 639

\bibitem[1997]{kob97} Kobulnicky, H.A., Skillman, E.D., Roy, J-R., Walsh, J.R., \& Rosa, M.R. 1997, ApJ, 477, 679

\bibitem[2003]{lee} Lee, H., McCall. M.L. \& Richer, M.G. 2003, AJ, 125, 2975

\bibitem[1997]{martina} Martin, C.L. 1997, ApJ, 491, 561

\bibitem[1986]{mathis} Mathis, J.S. 1986, ApJ, 301, 423

\bibitem[1985]{mccall} McCall, M.L., Rybski, P.M.,\& Shields, G.A. 1985, ApJS, 57, 1

\bibitem[1990]{moles} Moles, M., Aparicio, A., \& Masegosa, J. 1990, A\&A, 228, 310

\bibitem[1995]{olofsson} Olofsson, K. 1995, A\&AS, 111, 57 

\bibitem[1989]{ost89} Osterbrock, D.E. 1989, Astrophysics of Gaseous Nebulae and Active Galactic Nuclei, (University Science Books, Mill Valley, CA)

\bibitem[1999]{otte} Otte, B. \& Dettmar, R-J. 1999, A\&A, 343, 705

\bibitem[2002]{otteb} Otte, B., Gallagher, J.S.III, \& Reynolds, R.J. 2002, ApJ, 572, 823

\bibitem[1979]{pagel} Pagel, B.E.J., Edmunds, M.G., Blackwell, D.E., Chun, M.S., Smith, G. 1979, MNRAS, 189, 95

\bibitem[2000]{pilyugina} Pilyugin, L.S. 2000, A\&A, 362, 325

\bibitem[2001]{pilyuginb} Pilyugin, L.S. 2001, A\&A, 369, 594


\bibitem[1998] {rand} Rand, R.J. 1998, ApJ, 501, 137

\bibitem[1990]{randa} Rand, R.J., Kulkarni, S.R. \& Hester, J.J. 1990, ApJ, 352, 1

\bibitem[1989]{reynolds}Reynolds, R.J. 1983, ApJ, 268, 698

\bibitem[1989]{reynolds}Reynolds, R.J. 1989, ApJ, 345, 811

\bibitem[1979]{savage} Savage, B.D. \& Mathis, J.S. 1979, ARA\&A, 17, 73

\bibitem[1998]{schlegel} Schlegel, D.J., Finkbainer, D.P., \& Davis, M. 1998, ApJ, 500, 525 

\bibitem[2001]{shuster}Schuster, W.J., \& Parrao, L. 2001, RevMexA\&A, 37, 187

\bibitem[1993]{slavin} Slavin, J.D, Shull, J.M. \& Begelman, M.C. 1993, ApJ, 407, 83

\bibitem[1999]{stasinska} Stasinska, G. 1990, A\&A, 83, 501


\bibitem[1990]{strobel} Strobel, N.V., Hodge, P., \& Kennicutt, R.C. 1990, PASP, 102, 1241
\bibitem[1996]{vacca} Vacca, W.D., Garmany, C.D., \& Shull, J.M., 1996, ApJ, 460, 914
\bibitem[1989]{veilleux} Veilleux, S. \& Osterbrock, D.E. 1987, ApJS, 63, 295

\bibitem[2004]{wood} Wood, K., \& Mathis, J.S. 2004, MNRAS, 353, 1126
\end{thebibliography}
\end{document}